\documentclass[a4paper,fleqn,usenatbib]{mnras}

\usepackage[T1]{fontenc}
\usepackage{ae,aecompl}

\usepackage{graphicx}	
\usepackage{amsmath}	
\usepackage{amssymb}	

\title[Accurate parameters for HD 209458 and its planet]{Accurate parameters for HD 209458 and its planet from HST spectrophotometry}

\author[C. del Burgo and C. Allende Prieto]{
C. del Burgo$^{1}$\thanks{E-mail: cburgo@inaoep.mx}
and C. Allende Prieto$^{2,3}$
\\
$^{1}$Instituto Nacional de Astrof\'{\i}sica, \'Optica y Electr\'onica, Luis Enrique Erro 1, Sta. Ma. Tonantzintla, Puebla, Mexico\\
$^{2}$Instituto de Astrof\'{\i}sica de Canarias, 38205 La Laguna, Tenerife, Spain\\
$^{3}$Departamento de Astrof\'{\i}sica, Universidad de La Laguna, 38206 La Laguna, Tenerife, Spain
}

\date{Accepted XXX. Received YYY; in original form ZZZ}

\pubyear{2016}

\begin{document}
\label{firstpage}
\pagerange{\pageref{firstpage}--\pageref{lastpage}}
\maketitle

\begin{abstract}
We present updated 
parameters for the star HD 209458 and its transiting giant planet.
The stellar angular diameter $\theta$=0.2254$\pm$0.0017 mas is obtained from the average ratio between 
the absolute flux observed with the \textit{Hubble Space Telescope}
and that of the best-fitting Kurucz model atmosphere.
This angular diameter represents an improvement in precision of more than four times compared to available interferometric determinations.
The stellar radius $R_\star$=1.20$\pm$0.05~R$_{\sun}$ is ascertained by combining the angular diameter with the \textit{Hipparcos} trigonometric parallax, 
which is the main contributor to its uncertainty, and therefore the radius accuracy should be significantly improved with \textit{Gaia}'s measurements.
The radius of the exoplanet $R_\text{p}$=1.41$\pm$0.06~$R_\text{J}$ is derived from the corresponding transit depth 
in the light curve and our stellar radius.
From the model fitting, we accurately determine the effective temperature, $T_\text{eff}$=6071$\pm$20\,K, 
which is in perfect agreement with the value of 6070$\pm$24\,K 
calculated from the angular diameter and the integrated spectral energy distribution.
We also find precise values from recent Padova Isochrones, such as $R_\star$=1.20$\pm$0.06~R$_{\sun}$ 
and $T_\text{eff}$=6099$\pm$41\,K.
We arrive at a consistent picture from these methods and compare the results with those from the literature.
\end{abstract}




\section{Introduction}

Accurate stellar radii are required to determine accurate planetary radii from (spectro-)photometric transits.
Long-baseline optical/infrared interferometry provides the most direct way to ascertain stellar angular diameters, but it is limited to relatively nearby and bright stars 
\citep[e.g.:][]{vanbelle2009,baines2009,vonbraun2014,boyajian2015,ligi2015}.

Recently, \citet{allendeprieto2016} have shown that the comparison of absolute flux spectrophotometry of A stars 
from the STIS Next Generation Spectral Library \citep[NGSL;][]{gregg2006} with appropriate stellar atmosphere models leads to  
angular diameters that are more accurate than those from interferometry. 
Both methods require accurate parallaxes to finally achieve precise stellar radii.

Measurements of parallaxes from ground-based instruments are subject to distortions due to the Earth's atmosphere. 
Observations from space overcome this problem.
ESA's \textit{Hipparcos} was the first space mission fully dedicated to determine precise positions, distances, proper motions, 
and luminosities for more than 118,000 stars brighter than V=12.4 mag \citep[][]{ESA1997,vanleeuwen2007} over the entire sky. 
\textit{Hipparcos} has had a remarkable impact in many different areas of Astronomy. ESA's \textit{Gaia}, launched in 2013, is collecting 
astrometric, photometric, and spectroscopic data for about 10$^9$ stars with 3$\la$V$\la$20 mag\footnote{http://www.cosmos.esa.int/web/gaia}
with a much higher precision in position accuracy 
and much higher sensitivity than \textit{Hipparcos}.

HD 209458 is a nearby \citep[parallax $\Pi$=20.15$\pm$0.80~mas, ][]{vanleeuwen2007} G0V star 
with an estimated age of 4$\pm$2~Gyr \citep[][]{melo2006}.
It has been recently observed with the CHARA Array interferometer, finding that its limb-darkened angular diameter is  
$\theta_\text{LD}$= 0.225$\pm$0.007~mas \citep[][]{boyajian2015}.
It harbors HD 209458 b, the first transiting planet discovered \citep[][]{charbonneau2000}.
This is a short period inflated planet that has challenged planet formation theories \citep[][]{mardling2007}. 
HD 209458 is also one of only two systems with absolute stellar and planetary masses derived 
from high resolution spectroscopy \citep[][]{snellen2010}. The other is HD 189733 \citep[][]{dekok2013}.

Modeling the Spectral Energy Distribution (SED) of G-type stars is relatively easy in the
optical and infrared, where the continuum opacity is dominated by H$^-$ bound-free 
and free-free absorption, but becomes increasingly harder in the ultraviolet due to the
accumulation of spectral lines, the complex contribution of bound-free opacity from
neutral atoms, and the shift of the line formation from the photosphere to
higher atmospheric layers.

HD 209458 was observed with the instruments STIS and NICMOS onboard NASA/ESA \textit{Hubble Space Telescope} (HST). 
Observing from space avoids the scintillation noise resulting from air turbulence in the Earth's atmosphere, 
yielding an improvement of up to three orders of magnitude in (spectro-)photometric precision. 
Moreover, the limitations owing to the day-night cycle and variable weather conditions are eliminated.

In this paper we present results based on the analysis of the HST spectrophotometry of 
HD 209458\footnote{We adopted the nominal values of the International Astronomical Union (IAU) 2015 
Resolution B3 \citep[][]{prsa2016}: solar effective temperature of 5772\,K; 
GM$_{\sun}$=1.3271244 10$^{20}$ m$^3$ s$^{-2}$ and GM$_\text{J}$=1.2668653 10$^{17}$ m$^3$ s$^{-2}$, 
for the products of the gravitational constant by the masses of the Sun and Jupiter, respectively; 
R$_{\sun}$=6.957 10$^8$ m and $R_\text{J}$=7.1492 10$^7$ m for the Sun radius and 
the equatorial radius of Jupiter, respectively.
We applied the definition of the Resolution B2 of the XXVIII General Assembly of the IAU in 2012 
for the astronomical unit au=149 597 870 700 m.
We also employed G=6.67428 10$^{-11}$ m$^3$ kg$^{-1}$ s$^{-2}$, which is recommended 
by the IAU Working Group on Numerical Standards for Fundamental Astronomy, NSFA, \citep[][]{luzum2011}.
For the Stefan-Boltzmann constant, we adopted the CODATA 2014 value 
$\sigma$=(5.670367 $\pm$ 0.000013) $\times$ 10$^{-8}$ W~m$^{-2}$\,K$^{-4}$ \citep[][]{mohr2015}}.
Section \ref{obsmodspectra} describes the observed and modeled spectra, and the evolutionary models used here. 
In Section \ref{results}, the analysis and results on the properties of this host star and its planet are explained, 
including an accurate effective temperature for HD 209458 and the most accurate to date nearly model-independent 
radii for the star and its planet. This section also includes a comparison of our results with those in the literature. 
Section \ref{discussion} is devoted to a discussion on the potential precision and accuracy expected for 
the applied technique and a comparison with others. Finally, Section \ref{conclusions} presents our conclusions.

\section{Observed and modeled spectra}
\label{obsmodspectra}

\subsection{Absolute spectrophotometry}
\label{sec:abspec} 

The HST archive contains STIS CCD first order spectroscopy (G430L, G750M, and G750L) 
and NICMOS slitless grism imaging spectroscopy (G096, G141, and G206) 
for HD 209458.
The reduced data were downloaded\footnote{The FITS file \textit{hd209458\_stisnic\_006.fits}.} 
from the CALSPEC Calibration Database\footnote{http://www.stsci.edu/hst/observatory/crds/calspec.html}, 
which contains the composite stellar spectra for the flux standards on the HST system.

STIS data are fairly homogeneous, even though their wide spectral coverage relies on using different gratings.
The expected resolving power $R$\footnote{$R\equiv \frac{\lambda}{\delta\lambda}$, where $\delta\lambda$ is the 
full width at half maximum (FWHM) of a line spread function.} varies slightly with wavelength, 
as described in the STIS Instrument Handbook\footnote{http://www.stsci.edu/hst/stis/documents/handbooks/currentIHB/cover.html}.
According with the NICMOS Instrument Handbook\footnote{http://www.stsci.edu/hst/nicmos/documents/handbooks/handbooks/current\_NEW/nicmos\_ihb.pdf}
the resolving power is $R \sim$ 200 per pixel over the full field of view of the camera.
We found the FWHM values in the FITS header of the CALSPEC spectrum for HD 209458 to be smaller than
those needed to match this spectrum in the ultraviolet and visible (R$\approx$560-700). Therefore, we applied 
our own estimates of FWHM for each grating to properly smooth our theoretical models before comparing them to the observations.

\subsection{Stellar atmosphere models}
\label{models}

We computed synthetic spectra covering the wavelength range 0.2--2.5 $\mu$m 
with wavelength steps equivalent to 0.3 km s$^{-1}$. The calculations are based on Kurucz ATLAS9 model atmospheres 
\citep[][]{meszaros2012} and the synthesis code ASS$\epsilon$T  \citep[][]{koesterke2008,koesterke2009}, 
operated in 1D (plane-parallel geometry) mode. The reference solar abundances adopted in the model atmosphere and the 
synthesis are from \citet{asplund2005}.

The equation of state includes the first 92 
elements in the periodic table and 338 molecules \citep[][with some updates]{tsuji1964,tsuji1973,tsuji1976}. 
Partition functions are adopted from \citet{irwin1981}.
Bound-free absorption from H, H$^-$, HeI, HeII, and the first two ionization stages 
of C, N, O, Na, Mg,  Al, Si, 
Ca \citep[from the Opacity Project; see][]{cunto1993} 
and Fe \citep[from the Iron Project;][]{bautista1997, nahar1995} are included.
Line absorption is modeled in detail using the atomic and molecular 
(H$_2$, CH, C$_2$, CN, CO, NH, OH, MgH, SiH, and SiO) files 
compiled by Kurucz\footnote{kurucz.harvard.edu}.

Level dissolution near the Balmer series limit is accounted for \citep[][]{hubeny1994}.
The radiative transfer calculations include Rayleigh \citep[H;][]{lee2004} and electron 
(Thomson) scattering. The damping of H lines are treated in detail using Stark 
\citep[][]{stehle1994,stehle1999} and self-broadening 
\citep[for Balmer][for Lyman and Paschem lines]{barklem2000,ali1966}.

To ease the derivation of atmospheric parameters from the HST spectrophotometry, 
we computed a grid of model surface fluxes for effective 
temperatures in the range 5750 $\leq T_\text{eff} \leq$ 10,000\,K,
surface gravity 1.0 $\leq \log g \leq$ 5.0 ($g$ in cm s$^{-2}$), microturbulence -0.3 $\leq \log\xi_t \leq$ 0.9 ($\xi_t$ in km s$^{-1}$), 
metallicity -5 $\leq$ [Fe/H] $\leq$ +1.0, 
and $\alpha$-element enhancement -1.0 $\leq$ [$\alpha$/Fe] $\leq$ +1.0, 
in steps of 250\,K, 0.5 dex, 0.3 dex, 0.25 dex, and 0.25 dex, respectively. 
Spectra were convolved with a series of Gaussian kernels (see \S\ref{sec:abspec}) to
account for instrumental broadening. The effect of interstellar reddening was considered
following \citet{fitzpatrick1999}.

\subsection{Stellar evolution models}
\label{evolutionmodels}

We employed the {\sevensize PARSEC} (stellar tracks and isochrones with the PAdova \& TRieste Stellar Evolution Code) Isochrones \citep[version 1.2S:][]{bressan2012, chen2014, chen2015, tang2014}, with two different grids adequate to derive the stellar properties of the Sun and our target of interest. In the first grid, which is used to calibrate the system (given that HD 209458 is particularly similar to the Sun), the initial metallicity $Z_\text{ini}$ ranges from 0.0151 to 0.0211, in steps of 0.0005, age ($\tau$) from 4.1 to 5.3 Gyr, in steps of 0.01 Gyr, and initial mass ($M_\text{ini}$) ranging from 0.09 M$_{{\sun}}$ to the highest mass established by the stellar lifetime, in steps of 10$^{-4}$ M$_{{\sun}}$. In the second grid, $Z_\text{ini}$ goes from 0.0046 to 0.0246, in steps of 0.005, $\tau$ from 0.1 to 12.1 Gyr, in steps of 0.3 Gyr, and $M_\text{ini}$ spans from 0.09 M$_{{\sun}}$ to the highest mass given the stellar lifetime, in steps of 2 10$^{-3}$ M$_{{\sun}}$.

The {\sevensize PARSEC} models include the evolution of a star from its formation to the asymptotic giant branch phase for low- and intermediate mass stars, or the carbon ignition phase for massive stars. The age ruler by these models includes the pre-main sequence lifetime, which is about 40 Myr for the Sun and slightly 
lower for HD 209458.
These models provide, among other stellar parameters, actual mass, luminosity, effective temperature, surface gravity, bolometric magnitude, 
and magnitudes in a chosen photometric system (Johnson-Cousins for this research) as a function of the stellar age, metallicity, and the initial mass.
The radius can be trivially calculated from the mass and gravity, and the mean density from the mass and radius.
We only used the photometric bands $B$ and $V$ provided by {\sevensize PARSEC}, which were obtained with the responses and zero points 
derived by \citet{maizapellaniz2006}.
The corresponding stellar parameters for the regular sampling in M$_\text{ini}$ were computed from the tabulated values using linear interpolation.

\section{Analysis and results}
\label{results}

\subsection{Spectral analysis}
\label{ferre}

We searched for the values of $T_\text{eff}$, $\log g$, [Fe/H], [$\alpha$/Fe], and $\xi_t$,
associated with the model fluxes that best fit the spectrum of HD 209458 in 
the wavelength range from 0.29 to 2.5 $\mu$m.
For this fit we divided both observations and models, 
resampled on the same wavelength grid, by their own mean fluxes.
Interstellar extinction was considered as well, finding that the optimal solution led to 
a negative value ($E(B-V)$ = -0.09 mag), and therefore was neglected given the proximity 
of HD 209458 ($\sim$50 pc).
This is in agreement with the small value $E(B-V)$=0.003 mag found from the same HST data
by \citet{bohlin2014b}, who noted that the CALSPEC data for this star agrees with its 
preferred model to $\sim$2\%.

We defined three wavelength regions of interest: ultraviolet (UV: 0.29-0.50 $\mu$m), 
visible (VIS: 0.50-1.0 $\mu$m), and near-infrared (NIR: 1.0-2.5 $\mu$m). 
We performed fittings to VIS-NIR, VIS, UV-VIS-NIR, and UV-VIS, allowing all stellar parameters to vary (\textit{all/free}), 
fixing $\log g$ (\textit{g/fixed}), and fixing all parameters except $T_\text{eff}$ ($T_\text{eff}$/\textit{free}).
In this way, we could gauge the achieved accuracy and precision for determining stellar parameters for the different cases.

\begin{table}
	\centering
	\caption{Stellar parameters and 1$\sigma$ uncertainties of HD 209458 obtained from different methods: 1) BF:  Best-fitting Kurucz model atmosphere to ascertain $T_\text{eff}$, $\theta$ (through the comparison with the observed spectrum), and $R_\star$ (using $\theta$ and $\Pi$). $L_\star$ is then calculated from $R_\star$ and $T_\text{eff}$; 2) Procedure based on the bolometric flux to determine $T_\text{eff}$ using the theoretical model ($f_\text{bol}$-mod) and the observed spectrum ($f_\text{bol}$-obs) in combination with $\theta$. Also to obtain $L_\star$ using $\Pi$; 3) EM: Search for {\sevensize PARSEC} (version 1.2S) evolution models using the likelihood function ${\mathcal L}$; 4) Determination of $\log g$ from $R_\star$ (using BF) and either $\rho_\star$ derived from the light curve (BF-LC) or $M_\star$ derived from the {\sevensize PARSEC} models (BF-EM). BF-LC and BF-EM also permit to calculate $M_\star$ and $\rho_\star$, respectively. SynPhot: Synthetic photometry in the Johnson-Cousins $UBVRI$ and 2MASS systems extracted from the CALSPEC spectrum of this star.}
	\label{tab:stprop}
	\begin{tabular}{lcc}
		\hline
		Parameter                  &	Value$\pm$uncertainty    & Note   \\
		\hline
		$T_\text{eff}$ (K)         &	6071$\pm$20	    & BF; see \S\ref{ferre} \\
		$\shortparallel$           &	6070$\pm$24	    & $f_\text{bol}$-mod; see \S\ref{teffloggfeh} \\
		$\shortparallel$           &	6064$\pm$24	    & $f_\text{bol}$-obs; see \S\ref{teffloggfeh} \\
		$\shortparallel$           &	6099$\pm$41	    & EM; see \S\ref{isochrones} \\
		$\theta$ (mas)             &	0.2254$\pm$0.0017   & BF; see \S\ref{angdia} \\
		$R_\star$ (R$_{\sun}$)     &	1.20$\pm$0.05	    & BF; see \S\ref{stradii} \\
		$\shortparallel$           &	1.20$\pm$0.06	    & EM; see \S\ref{isochrones} \\ 
		$L_\star$ (L$_{\sun}$)     &	1.77$\pm$0.14	    & $f_\text{bol}$-mod; see \S\ref{luminosity}\\
		$\shortparallel$           &	1.76$\pm$0.14	    & $f_\text{bol}$-obs; see \S\ref{luminosity}\\
		$\shortparallel$           &	1.77$\pm$0.14	    & BF; see \S\ref{luminosity}\\
		$\shortparallel$           &	1.79$\pm$0.14	    & EM; see \S\ref{isochrones} \\
		$M_\text{bol}$ (mag)       &	4.11$\pm$0.09	    & EM; see \S\ref{isochrones} \\
		$\log g$ [cm~s$^{-2}$]     &	4.38$\pm$0.06	    & BF-LC; see \S\ref{star:mgd} \\
		$\shortparallel$           &	4.33$\pm$0.04	    & BF-EM; see \S\ref{star:mgd} \\
		$\shortparallel$           &	4.33$\pm$0.04	    & EM; see \S\ref{isochrones} \\
		$M_\star$ (M$_{\sun}$)     &	1.26$\pm$0.15	    & BF-LC; see \S\ref{star:mgd} \\ 
		$\shortparallel$           &	1.12$\pm$0.04	    & EM; see \S\ref{isochrones} \\ 
		$\rho_\star$ (g~cm$^{-3}$) &	0.91$\pm$0.11	    & BF-EM; see \S\ref{star:mgd} \\  
		$\shortparallel$           &	0.91$\pm$0.14       & EM; see \S\ref{isochrones} \\  
		$\tau$ (Gyr)               &	3.5$\pm$1.4	    & EM; see \S\ref{isochrones} \\
		\hline
		$U$                        &   8.235  & SynPhot; see \S\ref{photometry} \\
		$B$                        &   8.204  & $\shortparallel$ \\
		$V$                        &   7.655  & $\shortparallel$ \\
		$R$                        &   7.342  & $\shortparallel$ \\
		$I$                        &   7.037  & $\shortparallel$ \\
		$J$                        &   6.635  & $\shortparallel$ \\
		$H$                        &   6.346  & $\shortparallel$  \\
		$K_s$                      &   6.326  & $\shortparallel$ \\
		\hline
	\end{tabular}
\end{table}

The optimization was done using the FORTRAN90 code 
{\tt FER\reflectbox{R}E}\footnote{Available from http://hebe.as.utexas.edu/ferre} 
\citep[][]{allendeprieto2006} 
using  the Unconstrained Optimization BY Quadratic Approximation (UOBYQA) algorithm \citep[][]{powell2002}.
Model fluxes with any set of atmospheric parameters are derived by cubic Bezier interpolation 
\citep[see][and references therein]{auer2003} in the grid of model fluxes described in \S \ref{models}.

\begin{figure*}
\centering
 \includegraphics[width=175mm,angle=0]{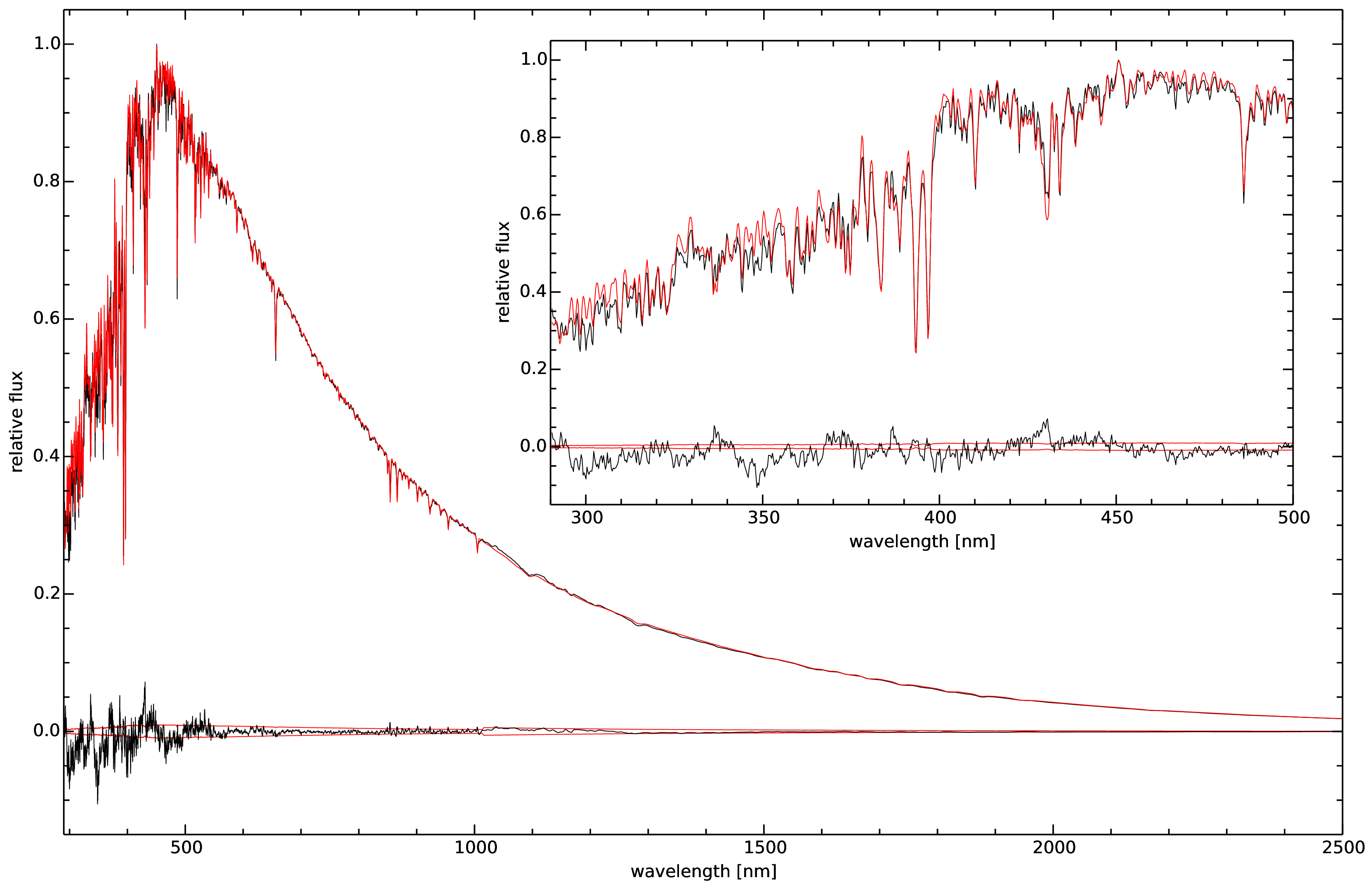}
 \caption{HST spectrum (solid black line) and best-fitting model (solid red line) for HD 209458.
 Also, at the bottom the residuals (black line) and the 1$\sigma$ uncertainties for the flux (red lines), stored in the FITS header of the CALSPEC spectrum. The spectra, residuals, and uncertainties between 290 and 500 nm are also zoomed.}
 \label{fig:fig1}
\end{figure*}

STIS spectra have been proven to be very useful to ascertain angular diameters of A-type stars, 
but a free parameter fitting does not necessarily provide accurate values of e.g., surface gravity \citep[][]{allendeprieto2016}. 
For HD 209458, the fitting with the lowest value for chi-square ($\chi^2$) was that of VIS-NIR-\textit{all/free}, 
but the extracted parameters ($T_\text{eff}$=6125\,K, $\log g$=4.78, [Fe/H]=-0.19) 
have extreme values (the highest ones for $T_\text{eff}$ and $\log g$, and the second lowest one of [Fe/H]
among all fittings).
Since our goal is to derive appropriate parameters of the star, 
a fitting was performed fixing all parameters\footnote{The value of $\log g$ was determined in 
\S\ref{star:mgd} and [Fe/H] is from \citet{torres2008}; see also \S\ref{isochrones} with very similar values.}
but $T_\text{eff}$, which must be well constrained given the broad wavelength range.

We indeed argue that VIS-$T_\text{eff}$/\textit{free} is our \textit{best} fitting ($T_\text{eff}$=6071\,K, $\log g$=4.38, 
[Fe/H]=0, [$\alpha$/Fe]=0, $\log \xi_t$=0.04) because: 
1) The VIS spectrum has a high signal-to-noise and many absorption lines that are very well modeled,
in contrast to the complex UV and the fainter nearly-featureless NIR; and
2) The zero-point of the flux scale is very well constrained in the visible (see \S\ref{photometry}).
We estimated an uncertainty in $T_\text{eff}$ of 20\,K, which is based on:
1) The differences between VIS-\textit{g/fixed} and VIS-NIR-\textit{g/fixed} (increase of $\approx$20\,K) and UV-VIS-NIR-\textit{g/fixed} (same value); and
2) The increment between VIS-$T_\text{eff}$/\textit{free} and VIS-NIR-$T_\text{eff}$/\textit{free} amounts to $\approx$15\,K, 
and the decrease with respect to UV-VIS-NIR-$T_\text{eff}$/\textit{free} is of $\approx$50\,K. 
The UV cannot be fitted as well as the optical, mainly due to the crowding of lines and continuum absorption by metals.

Therefore, our best fitting model (reduced chi-square $\chi^2_\text{red}$=0.462; N=1096 data points) 
to the CALSPEC spectrum relies on the VIS region,
arriving at $T_\text{eff}$=6071$\pm$20\,K (see Table \ref{tab:stprop}).
Fig. \ref{fig:fig1} shows the modeled and observed spectra.
We note the flux uncertainties (systematic and statistical) from the FITS header of the CALSPEC spectrum 
are overestimated by $\sim$50\% in the VIS region,
which is inferred from the comparison of these flux uncertainties with the residuals between the observed and modeled spectra.
This explains the low estimate of $\chi^2_\text{red}$.
In most of the UV the residuals are significantly larger than 
the estimated uncertainties stored in the FITS header of the HD 209458 spectrum, 
confirming the difficulties to model this region.
Conversely, the residuals in the NIR fall within the estimated flux uncertainties.

\subsection{Photometry}
\label{photometry}
We computed the Johnson-Cousins and 2MASS photometry from the CALSPEC spectrum of HD 209458 (see Table \ref{tab:stprop}). 
We adopted the Optimized Bessel Bandpass Functions presented by \citet{bohlin2015}. 
These correspond to the $UBVRI$ response curves of \citet{bessell2012} shifted by up to -31 \AA\ 
to minimize the differences between the Landolt photometry and the synthetic CALSPEC photometry
of 11 spectrophotometric flux standards.
The uncertainty of the monochromatic flux at 555.75 nm (555.6 nm in air) 
is 0.5\% or 0.005 mag \citep[see][and the CALSPEC Calibration Database]{bohlin2014b}.
Table 5 of \citet{bohlin2015} lists the magnitudes and uncertainties for the conversion to magnitudes of Vega on the Johnson-Cousing system. 
For example, $V$=0.028 mag, rms=0.005, and uncertainty in the mean of 0.003 mag.
These magnitudes for Vega were confirmed with our code\footnote{We used the same CALSPEC spectrum \textit{alpha\_lyr\_stis\_008.fits}, 
which combines IUE data from 115.2 to 167.5 nm, STIS CCD fluxes from 167.5 to 535.0 nm, 
and a Kurucz 9400 \,K model that matches very well the observed fluxes, but obviously lacks the noise in the observations, for wavelengths longer than 535.0 nm.}, 
with differences below 0.001 mag.

The values so obtained for HD 209458, $B$=8.204$\pm$0.010 and $V$=7.655$\pm$0.008 mag,
are slightly different from those computed using the response curves of \citet{maizapellaniz2006}, 
which lead to $B$ = 8.212 and $V$ = 7.649, but such differences are within the 1$\sigma$ uncertainties.
Our value of $V$ is similar to that of $V$=7.65 mag from \citet{bohlin2010} 
and slightly larger than that from \citet{bohlin2014b} of $V$=7.63\footnote{Note the last two values 
were not determined with the Optimized Bessel Bandpass Functions introduced by \citet{bohlin2015}.
Besides, the flux of Vega at 555.6 nm was updated to 3.44 $\times$ 10$^{-9}$ erg cm$^{-2}$ s$^{-1}$ $\AA^{-1}$ by \citet{bohlin2014a}.}.
It is also in agreement with the value of $V$=7.65$\pm$0.01 given by \citet{torres2008}.

The 2MASS $J$, $H$, and $K_s$ photometry was extracted from the CALSPEC spectrum  
using the response curves and zero points from \citet{cohen2003a}.
Our values of $J$=6.635, $H$=6.346, and $K_s$=6.326 are in fair agreement with those of 
$J$=6.591$\pm$0.020, $H$=6.37$\pm$0.04, and $K_s$=6.308$\pm$0.026 from \citet{cutri2003}.
Given the uncertainties in the 2MASS passbands, absolute values of differences correspond to 
$\sim$2$\sigma$, $\sim$0.5$\sigma$, and $\sim$0.5$\sigma$, respectively.

There are different response curves for the Johnson and Cousins systems in the literature, yielding different 
synthetic magnitudes extracted from a certain model spectrum. Apart from the above-mentioned Optimized Bessel Bandpass Functions,
we have employed the response curves from \citet{cohen2003b}, \citet{maizapellaniz2006}, and \citet{mann2015} to compute the magnitudes of 
HD 209458 in the $UBVRI$ filter-bands\footnote{\citet{maizapellaniz2006} provides $U$, $B$, and $V$ response curves.}. 
The differences between the maximum and minimum extracted values are 0.09, 0.016, 0.02, 0.03, and 0.013 mag for $U$, $B$, $V$, $R$, and $I$, respectively. 
The most extreme differences are 0.09 mag in the $U$ band.  
Note that bandpass functions may be multiplied by a mean atmospheric transmission spectrum appropriate 
for a certain observing site \citep[e.g.,][for CTIO, Chile]{cohen2003b}.
Earth's atmospheric transmission variations lead to systematic errors in the UV.
This has an impact on the results from model atmosphere fittings to broad-band photometry, 
together with the reduced number of points used to constrain the solution. 
Our technique overcomes this problem, since it uses the high signal-to-noise CALSPEC spectrum, 
which is absolutely calibrated in flux and presents $\simeq$550 resolved elements in the wavelength range between 500 and 1000 nm
($\simeq$1030 resolved elements in the full wavelength range of STIS and NICMOS).

\subsection{Stellar angular diameter}
\label{angdia}
\begin{figure}
\centering
 \includegraphics[width=85mm,angle=0]{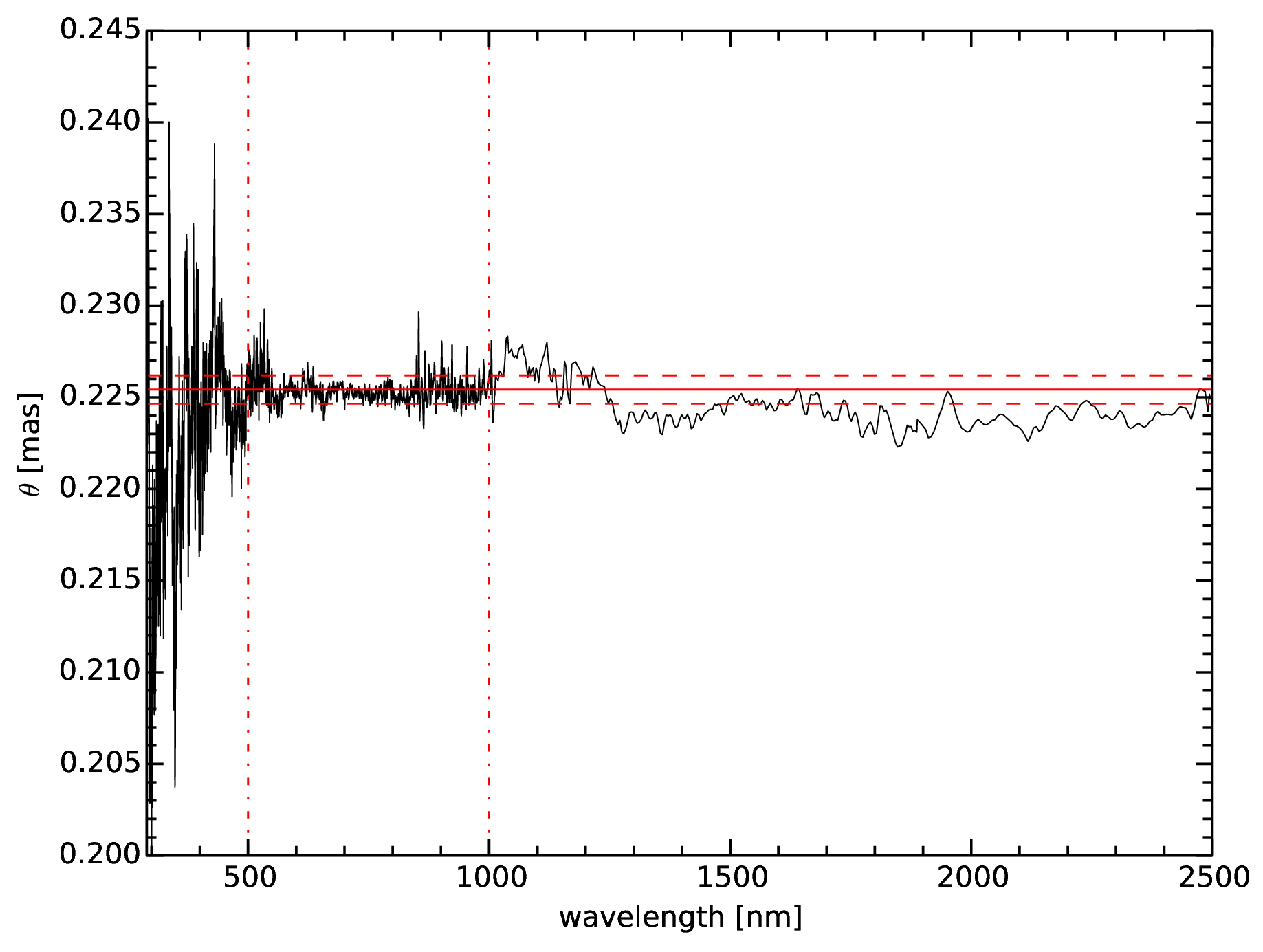}
 \caption{Angular diameter $\theta \simeq 2 \sqrt{f/F}$ versus wavelength. The vertical dash-dotted lines indicate the wavelength limits used to obtain the mean value of $\theta$ and its standard deviation, which are marked by the continuous line and dashed lines, respectively. Note that although the fitting is applied to the VIS region, the match is very good in the full wavelength range. The total uncertainty in $\theta$ was calculated quadratically adding to the above-mentioned standard deviation the contributions from the error in the zero-point on the flux scale and the error in the spectral shape due to the uncertainties in the stellar parameters (see \S \ref{angdia}).}
 \label{fig:fig2}
\end{figure}

In a similar way to \citet{delburgo2010} and \citet{allendeprieto2016}
the angular diameter $\theta \simeq 2 R/d = 2 \sqrt{f/F}$ for HD 209458 was determined from
the average ratio $<f/F>$ between the flux observed and that predicted by the model at the stellar surface in the 0.5-1.0 $\mu$m wavelength range. 
We used the best fitting model obtained in \S\ref{ferre} ($T_\text{eff}$=6071 \,K, $\log g$= 4.38, [Fe/H]=0, [$\alpha$/Fe]=0, $\log \xi_t$=0.04)
and find the angular diameter to be $\theta$=0.2254$\pm$0.0017 mas (see Table \ref{tab:stprop}).
Our value is more precise than the interferometric one of 0.225$\pm$0.007 mas from \citet{boyajian2015}.
It is also better than the value of 0.224$\pm$0.004 mas determined by \citet{casagrande2010} with a methodology similar to ours,
probably due to the use of approximate SEDs provided with the model atmospheres adopted in their analysis.
Fig. \ref{fig:fig2} shows $\theta$ versus wavelength, where it is observed that the fitting based on the VIS region is very good in the full spectral range.
Indeed, we found a very similar value for the stellar angular diameter in the wavelength ranges VIS and VIS-NIR.

The uncertainty $\sigma(\theta)$ is quantified by adding in quadrature three terms:
1) a random error contribution estimated from the scatter of the flux ratio across the selected wavelength interval (0.0008 mas);
2) the error in the spectral shape because of the uncertainties in the stellar parameters, which is computed from a MonteCarlo approach with 100 trials (0.0014 mas); and 
3) the error contribution of 0.25\% that corresponds to propagating the uncertainty of the flux scale of 0.5\% in the zero-point (0.0006 mas).

Our result is largely model-independent, since the impact of the uncertainties in the parameters describing the model is moderately low. 
The uncertainties introduced to derive this component of $\sigma(\theta)$ are 
$\sigma(T_\text{eff})$=20\,K, $\sigma(\log g)$=0.06 dex, and $\sigma$([Fe/H])=0.05 dex.
We found $\sigma(T_\text{eff})$ to be the major contributor, while the effect of $\sigma(\log g)$ is negligible. 

For A-type stars, we found that the values of $\log g$ and [Fe/H] derived 
from spectral fitting depend heavily on the Balmer and Paschen jumps, and the near-UV line absorption, respectively \citep[][]{allendeprieto2016}. 
In that study some of the extracted values for $\log g$ were lower (by up to 0.5 dex) than those ascertained 
from stellar structure models.
However, metallicity and surface gravity (as well as the effective temperature) have a
very limited impact on the spectral energy distribution of A-type stars in the 
wavelength range 0.40-0.80 $\mu$m \citep[see also][]{delburgo2010}. 
Our analysis for the G-type star HD 209458 relies on an accurate determination of $T_\text{eff}$ and $\log g$,
which is fixed in our fitting. This leads to a slightly higher value of $\sigma(\theta)$ compared to that 
derived from an analysis in which all parameters are free to vary.

\subsection{Stellar radius}
\label{stradii}
Combining the angular diameter obtained in the previous section with the trigonometric parallax 
from \textit{Hipparcos} \citep[$\Pi$=20.15$\pm$0.80~mas, ][]{vanleeuwen2007} for HD 209458 
we derive the stellar radius $R_\star$=1.20 $\pm$ 0.05~R$_{\sun}$ (see Table \ref{tab:stprop}).
This value is in excellent agreement with that from \citet{boyajian2015} (1.20$\pm$0.06~R$_{\sun}$), 
obtained from their limb-darkened angular diameter and the \textit{Hipparcos} parallax. 
But we also found our result to be consistent with others from stellar evolution models such as those from 
\citet{bonfanti2015} (1.20$\pm$0.04 R$_{\sun}$), \citet{torres2008} (1.155$^{+0.014}_{-0.016}$~R$_{\sun}$),
and our own analysis described in \S\ref{isochrones} (1.20$\pm$0.06 R$_{\sun}$).
Note our precision is similar to that of \citet{bonfanti2015}, who employed {\sevensize PARSEC} \citep[version 1.0, ][]{bressan2012},
but quite different to that of \citet{torres2008},
who used Yonsei-Yale (Y$^2$) series by \citet{yi2001} constrained by [Fe/H], $T_\text{eff}$, and $\frac{a}{R_\star}$ 
($a$ is the semi-major axis of the orbit) using a likelihood function.

\subsection{Bolometric flux and effective temperature}
\label{teffloggfeh}

We determined the bolometric flux $f_\text{bol}$ = (2.289$\pm$0.011) $\times$ 10$^{-8}$ erg cm$^{-2}$ s$^{-1}$ from the 
integration between 0.29 and 300 $\mu$m of the HD 209458 spectrum, extended towards the UV ($\lambda <$ 0.29 $\mu$m) 
with a Kurucz model with $T_\text{eff}$=6071 \,K. The fluxes beyond 2.5 $\mu$m come from a Kurucz model 
introduced to complete the SED \citep[][]{bohlin2010}. 
The ultraviolet and infrared portions added to the SED respectively contribute only 1.6\% and 2.9\% of the total flux. 
The use of different models (with other stellar parameters values within the uncertainties) to complete the observed 
STIS+NICMOS spectrum has a negligible effect on the results presented here. 

Alternatively, we obtained $f_\text{bol}$= (2.298$\pm$0.011) $\times$ 10$^{-8}$ erg cm$^{-2}$ s$^{-1}$ from the best model 
derived in \S\ref{ferre}. Any of our values of $f_\text{bol}$ (which are consistent with each other) is $\sim$5 times 
more precise than that of \citet{boyajian2015} with $f_\text{bol}$= (2.33$\pm$0.05) $\times$ 10$^{-8}$ erg cm$^{-2}$ s$^{-1}$, 
which they derived from a model fitted to broad-band photometry collected from the literature.
\citet{casagrande2010} arrived at $f_\text{bol}$= (2.335$\pm$0.025) $\times$ 10$^{-8}$ erg cm$^{-2}$ s$^{-1}$,
adopting [Fe/H] = 0.03$\pm$0.02, $\log g$=4.50$\pm$0.04, Tycho 2, and 2MASS photometry.

From the bolometric flux we determine the effective temperature 
($T_\text{eff}=\left(\frac{4 f_\text{bol}}{\sigma \theta^2}\right)^{0.25}$, where $\sigma$ is the Stefan-Boltzmann constant). 
We have thus found the effective temperature to be $T_\text{eff}$=6064$\pm$24\,K from the CALSPEC spectrum, and
6070$\pm$24\,K from the best-fitting model, in consonance with that derived in \S\ref{ferre}. These results 
(compiled in Table \ref{tab:stprop}) demonstrate the high precision of the spectral shape and flux zero-point of the HST data, 
as well as the excellent match to our best-fitting stellar atmosphere model.
\citet{boyajian2015} found that $T_\text{eff}$=6092$\pm$103\,K from their $f_\text{bol}$ and interferometric limb-darkened 
angular diameter, and \citet{casagrande2010} obtained $T_\text{eff}$=6113$\pm$49\,K. 
Our more precise $f_\text{bol}$ and $\theta$ (see \S\ref{angdia}) lead to a significantly 
more precise value for the effective temperature.

\subsection{Stellar luminosity}
\label{luminosity}
First, we determined the stellar luminosity from the bolometric flux and the \textit{Hipparcos} parallax (L $\propto \frac{f_\text{bol}}{\Pi^2}$), 
which leads to $L_\star$=1.76$\pm$0.14~L$_{\sun}$ and 1.77$\pm$0.14~L$_{\sun}$ using the values of $f_\text{bol}$ 
obtained from the observed and theoretical spectrum, respectively (see \S\ref{teffloggfeh}). 
Note $\sigma(L_\star)$ is dominated by the uncertainty in the parallax. 
Alternatively, it is found that $L/L_{{\sun}}=\left(R/R_{\sun}\right)^2 \left(T_\text{eff}/T_{\text{eff},{\sun}}\right)^4$, 
yielding $L_\star$=1.77$\pm$0.14~L$_{\sun}$ from $R_\star$ (see \S\ref{stradii}) and $T_\text{eff}$ (see \S\ref{ferre}).
These values are compiled in Table \ref{tab:stprop}.

Our results are in good agreement with the values of $L_\star$=1.79$\pm$0.15~L$_{\sun}$ and $L_\star$=1.77$\pm$0.01~L$_{\sun}$ 
from \citet{boyajian2015} and \citet{bonfanti2015}, respectively. 
The former is obtained from a model atmosphere fitting to broad-band photometry and the latter is based on stellar evolutionary models.
Our value of $L_\star$=1.79$\pm$0.14~L$_{\sun}$ derived from {\sevensize PARSEC} models (see \S\ref{isochrones}) is also consistent with all these values.
Note, however, the very small uncertainty in the luminosity given by \citet{bonfanti2015} even though they used Padova Isochrones (version 1.0).
Also from stellar evolutionary models, \citet{torres2008} found a relatively low value of $L_\star$=1.62$\pm$0.10~L$_{\sun}$, 
but it is still in agreement with the others.

\subsection{Stellar mass, mean density and gravity}
\label{star:mgd}

We applied two different approaches to determine two of three stellar parameters (mean density, mass, and gravity)
from the stellar radius (see \S\ref{stradii}) combined with either the mass or the mean density from the literature 
or from the {\sevensize PARSEC} models (see \S\ref{isochrones}). 
First, given the radius and the mass, we found the mean stellar density to be $\rho=\frac{3M}{4\pi R^3}$ and 
gravity $g=\frac{GM}{R^2}$, where $G$ is the gravitational constant.
There is good agreement among results for the stellar mass $M_\star$=1.11$\pm$0.02~M$_{\sun}$ \citep[][]{bonfanti2015}, 
1.12$\pm$0.03~M$_{\sun}$ \citep[][]{torres2008}, and 1.12$\pm$0.04~M$_{\sun}$ (see \S\ref{isochrones}).
We used our values of $M_\star$ (see \S\ref{isochrones}) and $R_\star$ 
(see \S\ref{stradii}) to arrive at $\rho_\star$=0.91$\pm$0.11 g cm$^{-3}$ 
and $\log g$= 4.33$\pm$0.04 (with $g$ in cm~s$^{-2}$) (see Table \ref{tab:stprop}).

Second, we used our stellar radius and a value for the mean stellar density ascertained 
by \citet{torres2008} from the (high signal-to-noise) HST/STIS transit light curve \citep[][]{brown2001}. 
\citet{torres2008} determined $\rho_\star$=1.024$\pm$0.014~g~cm$^{-3}$ from the expression 
$\rho_\star=\frac{3\pi}{GP^2}\left(\frac{a}{R_\star}\right)^3 - \rho_\text{p}\left(\frac{R_\text{p}}{R_\star}\right)^3$,
inserting the semi-major to stellar radius ratio $\frac{a}{R_\star}$ and the period $P$.
They neglected the second term, which is generally very small because the planetary radius $R_\text{p}\ll{R_\star}$. 
From this mean stellar density and our $R_\star$ we calculated $M_\star$=1.26$\pm$0.15~M$_{\sun}$ and more significantly 
$\log g$=4.38$\pm$0.06 ($g$ in cm~s$^{-2}$) (see Table \ref{tab:stprop}).
The estimated value for the uncertainty of $\log g$ derived in \S\ref{isochrones} is smaller, 
but we argue that $\log g$=4.38$\pm$0.06 (with $g$ in cm~s$^{-2}$) is more accurate and nearly model-independent, 
so we decided to use it as a fixed parameter in the fitting process described in \S\ref{ferre}.

\subsection{Isochrones}
\label{isochrones}

\begin{table}
	\centering
	\caption{Stellar parameters ($T_\text{eff}$, $\log g$, $\tau$, $L$, $M_\text{bol}$, $M$, $R$, and $\rho$) and 1$\sigma$ uncertainties 
	for the Sun inferred from our method on the {\sevensize PARSEC} models for the initial solar metallicity from \citet{bressan2012}, 
	$M_\text{V}$=4.862$\pm$0.020 from \citet{pecaut2013} and $B-V$=0.653$\pm$0.003 from \citet{ramirez2012}. 
	Corrections have been applied to be consistent with the IAU 2015 Resolution B3 nominal values 
	for R$_{\sun}$, M$_{\sun}$, and L$_{\sun}$ \citep[][]{prsa2016}.}
	\label{tab:stpropevolution}
	\begin{tabular}{lccc}
		\hline
		Parameter                    & Value$\pm$uncertainty \\
		\hline
		$T_\text{eff}$ (K)           & 5778$\pm$8         \\
		$\log g$  [cm~s$^{-2}$]      & 4.435$\pm$0.006    \\
		$\tau$  (Gyr)                & 4.7$\pm$0.3        \\
		$L$/L$_{\sun}$               & 1.000$\pm$0.014    \\
		$M_\text{bol}$ (mag)         & 4.740$\pm$0.015    \\
		$M$/M$_{\sun}$               & 0.998$\pm$0.004    \\
		$R$/R$_{\sun}$               & 1.002$\pm$0.007    \\
		$\rho$/$\rho_{\sun}$         & 0.992$\pm$0.022    \\
		\hline
	\end{tabular}
\end{table}

In order to derive the stellar parameters from {\sevensize PARSEC} models, we followed a procedure similar to that described by \citet{jorgensen2005}, 
but using $B-V$ instead of $T_\text{eff}$, and with flat priors. 
We estimated the different parameters from the likelihood function, ${\mathcal L}$, given in their Equation 4, and integrating that over the model parameters, 
namely age ($\tau$), initial mass ($M_\text{ini}$), and initial metallicity ($Z_\text{ini}$). 
${\mathcal L}$ equals the probability of getting the observed data ($M_\text{V}$, $B-V$, and metallicity) 
for a given set of parameters ($\tau$, $Z_\text{ini}$, and $M_\text{ini}$). 
For $T_\text{eff}$, for instance, the estimated value and its variance are:

\begin{equation}
E(T_\text{eff})        = \int ({\mathcal L} \times T_\text{eff})~dM_\text{ini}~d\tau~dZ_\text{ini}
\end{equation}
\begin{equation}
Var(T_\text{eff})      = \int ({\mathcal L} \times [T_\text{eff} - E(T_\text{eff})]^2)~dM_\text{ini}~d\tau~dZ_\text{ini}
\end{equation}

We employed the solar parameters to calibrate our method with the grid mentioned in \S\ref{evolutionmodels}.
We adopted for the Sun the values $M_\text{V}$= 4.862$\pm$0.020 mag from \citet{pecaut2013} 
and $B-V$=0.653$\pm$0.003 mag from \citet{ramirez2012}, 
together with the initial solar metallicity $Z_\text{ini}$=0.01774 given by \citet{bressan2012}. 
The modeled values for the Sun in the corresponding {\sevensize PARSEC} isochrone are $M_\text{V}$= 4.841 mag, $M_\text{B}$=5.530 mag, 
and $M_\text{bol}$=4.769 mag.
Therefore, we had to introduce corrections to the model variables $M_\text{B}$ (subtracting 0.015 mag), 
and $M_\text{V}$ (adding 0.021 mag), in order to be consistent with the magnitudes from \citet{pecaut2013} and \citet{ramirez2012}.
We also applied a correction to $M_\text{bol}$ (subtracting 0.03 mag) to be in accordance with 
the IAU 2015 Resolution B2 value of 4.74 mag\footnote{http://www.iau.org/news/announcements/detail/ann15023/}.
If these corrections are not made then the output values are significantly different to the Sun's values, in particular $T_\text{eff}$=5837\,K. 
In addition, we applied corrections to be consistent with the IAU 2015 Resolution B3, 
on recommended nominal constants \citep[][]{prsa2016}.
These corrections renormalize the inferred values from {\sevensize PARSEC} to match the adopted ones 
for the Sun's parameters at the IAU 2015. The corrections for
M$_{\sun}$\footnote{The Sun's mass adopted in {\sevensize PARSEC} was provided by Alessandro Bressan (private communication).}, 
L$_{\sun}$, and R$_{\sun}$ are respectively of 0.034\%, 0.47\%, and 0.040\%. The
results are given in Table \ref{tab:stpropevolution}.
Note the good agreement between our inferred values and those approved by the IAU,
e.g. the effective temperature $T_\text{eff}$=5772\,K. 

To determine the stellar parameters of HD 209458 using {\sevensize PARSEC} Isochrones and the grid introduced 
in \S\ref{evolutionmodels}, we used as input values the metallicity 
[Fe/H]=0.00$\pm$0.05 dex\footnote{This leads to a solar value for $Z_\text{ini}$, taking into account the 
expected evolution of the metallicity for a star similar to the Sun \citep[see][]{bonfanti2015}.} 
of \citet{torres2008}, 
and our values $M_\text{V}$= 4.18$\pm$0.09 mag and $B-V$= 0.549$\pm$0.013 mag, which were calculated from 
the \textit{Hipparcos} parallax and the values of B and V obtained in \S \ref{photometry}.
The values of the stellar properties derived from this analysis 
($T_\text{eff}$, $R_\star$, $L_\star$, $M_\text{bol}$, $\log g$, $M_\star$, $\rho_\star$, and $\tau$) 
are given in Table \ref{tab:stprop}. 
We found negligible differences in the inferred stellar parameters when using steps in age of 0.1, 0.2, and 0.4 Gyr instead of 0.3 Gyr
, as well as for steps in $Z_\text{ini}$ as low as 0.0005 instead of 0.005.
Also when considering different steps (as low as 0.0001 M$_{\sun}$) in $M_\text{ini}$, which were generated from a linear interpolation.

\citet{bonfanti2015} also employed Padova Isochrones (v1.0) finding that HD 209458 has 
$T_\text{eff}$=6084$\pm$63 $K$ and $\log g$=4.30$\pm$0.10 (with $g$ in cm~s$^{-2}$),
which are in good agreement with our values 
$T_\text{eff}$=6099$\pm$41 $K$ and $\log g$=4.33$\pm$0.04 (with $g$ in cm~s$^{-2}$).
We note that we could have subtracted 6\,K to the inferred effective temperature since this is the formal 
difference between the output and input values for the Sun (see Table \ref{tab:stpropevolution}), but we did 
not do it since the difference is within the uncertainties.
\citet{bohlin2014b} used the same HST data employed in this analysis and Kurucz models
to arrive at $T_\text{eff}$=6100 \,K, $\log g$=4.20 ($g$ in cm~s$^{-2}$), $[\frac{M}{H}]$=-0.04, and $E(B-V)$=0.003 mag.
From an spectroscopic analysis \citet[][]{santos2004} found $T_\text{eff}$=6117$\pm$26 $K$, 
$\log g$= 4.48$\pm$0.08 ($g$ in cm~s$^{-2}$), [Fe/H]=0.02$\pm$0.03, and $\xi_t$=1.40$\pm$0.06 km s$^{-1}$. 
Their value for the effective temperature is higher than ours, 6071$\pm$20 $K$ (see \S\ref{ferre}).
\citet{torres2008} derived T$_\text{eff}$=6065$\pm$50\,K, $\log g$=4.42$\pm$0.04 ($g$ in cm~s$^{-2}$), and [Fe/H]=0.00$\pm$0.05 dex,
from a combination of different values in the literature. 
They also obtained $\log g$=4.361$^{+0.007}_{-0.008}$ ($g$ in cm~s$^{-2}$) from stellar evolutionary models.
Note the very small estimated uncertainties for the last value. In general, all these values are consistent with each other.
Our derived age $\tau$=3.5$\pm$1.4 Gyr is also in good agreement with other values in the literature, such as 
3.1$^{+0.8}_{-0.7}$~Gyr \citep[][]{torres2008}, 4.0$\pm$1.2~Gyr \citep[][]{bonfanti2015}, 
and 4$\pm$2~Gyr \citep[][]{melo2006}.
See also \S\ref{stradii} and \S\ref{star:mgd} for the comparisons of $R_\star$,  $M_\star$, and $\rho_\star$.

\subsection{Planetary radius and semi-major axis}
\label{plrad}

We determined the radius of HD 209458b from the radius of the host star and the transit depth.
The latter provides the planet to star radius ratio. \citet{torres2008} found that 
$\frac{R_\text{p}}{R_\star}$= 0.12086$\pm$0.00010 and $\frac{a}{R_\star}$=8.76$\pm$0.04
from the STIS light curve \citep[][]{brown2001}.

In general, the radii ratios from the \textit{Spitzer}/IRAC light curves are much less affected by limb-darkening effects and the presence of spots.
Thus, we have also considered the ratio $\frac{R_\text{p}}{R_\star}$= 0.12099$\pm$0.00029 of \citet{evans2015}
to determine $R_\text{p}$.
They noted that its effective radius is constant or modestly decreasing from 4.5 to 8 $\mu$m.

We used our $R_\star$ ascertained in \S\ref{stradii}. Our results for the radius and semi-major axis of this exoplanet are presented in Table~\ref{tab:plprop}.
Note the values of $R_\text{p}$ obtained from the two ratios (1.41$\pm$0.06 and 1.42$\pm$0.06 $R_\text{J}$) are consistent with each other.

\begin{table}
	\centering
	\caption{Planetary and orbital properties and corresponding 1$\sigma$ uncertainties of HD209458b: radius, mass, mean density, gravity, and semimajor axis. 
	We used our $R_\star$ obtained in \S\ref{stradii} and the ratios $\frac{R_\text{p}}{R_\star}$ from \citet{torres2008} (R\&rT) 
	and \citet{evans2015} (R\&rE) to determine $R_\text{p}$. R\&gT: Same radius and $\log g_\text{p}$ from \citet{torres2008} to determine the mass.
	$\rho_\text{p}$ was obtained from $R_\text{p}$ and $M_\text{p}$ (M\&R).
	R\&aT: $a$ was ascertained from the ratio $\frac{a}{R_\star}$ from \citet{torres2008} and $R_\star$.}
	\label{tab:plprop}
	\begin{tabular}{lcc}
		\hline
		Parameter                              &   Value$\pm$uncertainty                 &     Note    \\
		\hline
		$R_\text{p}$ ($R_\text{J}$)            &	    1.41$\pm$0.06                &     R\&rT; see \S\ref{plrad} \\
		$\shortparallel$                       &	    1.42$\pm$0.06                &     R\&rE; see \S\ref{plrad} \\
		$M_\text{p}$ ($M_\text{J}$)            &	    0.74$\pm$0.06	  	 &     R\&gT; see \S\ref{planets:mgd} \\
		$\rho_\text{p}$ (gr cm$^{-3}$)         &	    0.32$\pm$0.05	         &     M\&R; see \S\ref{planets:mgd} \\
		$\log g_\text{p}$ [cm~s$^{-2}$]        &	  2.963$\pm$0.005	  	 &     \citet{torres2008}    \\
		$a$ (au)                               &	    0.0490$\pm$0.0020	         &     R\&aT;  see \S\ref{planets:mgd} \\
		\hline
	\end{tabular}
\end{table}

\subsection{Planetary mass, mean density, and gravity}
\label{planets:mgd}

The surface gravity $\log g_\text{p}$=2.963$\pm$0.005 (with g cm~s$^{-2}$) of the planet HD 209458b was determined by \citet{torres2008} 
from the semi-amplitude of the radial velocity and the transit light curve.
We found the planetary mass to be $M_\text{p}$ = 0.74$\pm$0.06 $M_\text{J}$ from $\log g_\text{p}$ and $R_\star$ (see \S\ref{stradii}). 

\citet{snellen2010} measured spectral lines originating from HD 209458b, and determined the masses of the planet and the star 
from the radial velocity semi-amplitudes of both objects. They obtained $M_{\star}$ = 1.00$\pm$0.22 $M_{{\sun}}$ and $M_\text{p}$ = 0.64$\pm$0.09 $M_\text{J}$.

We have also determined the mean density of the exoplanet, just from the mass and the radius.
Table~\ref{tab:plprop} shows the values found for the mass, mean density, and surface gravity of HD 209458b.

\section{Discussion}
\label{discussion}
We have ascertained the effective temperature and angular diameter of the G-type star HD 209458
from the comparison of absolute flux spectra obtained with HST and Kurucz stellar atmosphere models.
The high accuracy in $\theta$ propagates to other stellar parameters, such as the linear radius (whose uncertainty is dominated by 
that in the parallax) and the surface gravity, computed adopting a precise value of the mean density from 
the HST/STIS transit light curve \citep[][]{torres2008}. The determination of the planetary parameters also benefits from the high
accuracy achieved for the stellar parameters.

Our determination of the angular diameter $\theta$ is more than four times more precise than that from 
interferometry by \citet{boyajian2015}. These authors warned about their angular size for HD 209458 being at 
the resolution limit of CHARA/PAVO and that they used calibrators up to $\sim$30\% smaller than this object. 
Nevertheless, their value is consistent with ours.

While interferometry can be only applied to bright nearby stars, absolute flux spectrophotometry compared with 
stellar atmosphere models yields very precise angular diameters for a wide range of luminosities. 
\citet{allendeprieto2016} have shown that while for a bright A star such as Vega the achieved uncertainty 
of 0.4\% in $\theta$ is similar to that from interferometry, for fainter A stars (V$\sim$3-6 mag) the precisions from 
the absolute flux comparison method are several times better than the respective interferometric precisions.

For the faintest A stars in the sample of \citet{allendeprieto2016} (7.1 $<V<$ 7.5 mag; $d>$ 100 $pc$) the uncertainties 
in the linear stellar radii derived from spectrophotometry are $\sim$10\%.
These are derived from the uncertainties $\sigma(\theta)$ ($<$0.7\%) and $\sigma(\Pi)$ ($\sim$10\%). 
For HD 209458 we arrived at $\sigma(\theta)$ = 0.75\% and $\sigma(\Pi)$ = 4\%.
Therefore, the precision in all these stellar radii is limited by the uncertainty in the \textit{Hipparcos} parallax. 
These radii could be recalculated with much higher precision from the forthcoming \textit{Gaia} parallaxes ($\sigma(\Pi)\sim$ 0.05\%),
yielding stellar radius uncertainties under 1\%. 

For fast-rotating A-type stars interferometry offers the advantage of directly gauging the oblateness and gravity darkening, 
but this is generally only possible for nearby bright stars. Just recently, some advances have been performed to correct 
for the effect of rotational distortion in fainter A stars and to properly quantify their sizes \citep{jones2015}.

Stellar evolution models such as those used in \citet{torres2008} and the ones employed to this research can reproduce 
well the stellar properties of HD 209458 and other sun-like stars. But they may present difficulties to predict the properties 
of more massive stars and low mass stars. For example, from the comparison of interferometric angular diameters with 
stellar evolutionary models for a sample of K-and M-dwarfs \citet{boyajian2012} found that such models overestimate 
the effective temperatures of stars with $T_\text{eff}$ < 5000\,K by $\sim$3\% and underestimate their radii for 
$R_\star<$0.7 R$_{\sun}$ by $\sim$5\%.

Asteroseismology is an alternative technique to ascertain fundamental properties of pulsating stars. 
Recently, it has been successfully applied to derive linear stellar radii and masses with typical uncertainties of $\sim$3\% 
and $\sim$7\%, respectively, for a sample of 66 planet-candidate host stars observed 
with Kepler \citep[7.8 $<K_\text{p}<$ 13.8 mag;][]{huber2013}. 
Space asteroseismology is usually combined with ground-based high-resolution spectroscopy values of 
$T_\text{eff}$ and [Fe/H] to precisely obtain the stellar parameters of sun-like and red giants.
The stellar radius and surface gravity can be tightly constrained by a precise value of the effective temperature.
While the metallicity permits to constrain the stellar mass and age \citep[see review of][and references therein]{chaplin2013}.
Mean density, surface gravity and radius determined from this technique are shown to be largely model independent.

\citet{huber2013} found, for more than half of their sample, discrepancies greater than 50\% between the 
densities derived from asteroseismology and those from transit models assuming circular orbits (mostly underestimated). 
They concluded this was due to systematics in the modeled impact parameters 
or the presence of candidates in eccentric orbits.
It will be eventually possible to compare, for an statistically significant sample of stars, the values 
of their radii obtained from all the above-mentioned techniques with uncertainties better than 1\%. 
This will allow us to further search for and reduce systematic errors.

It is well-known that stellar activity (e.g., spots) may affect the determination of the planetary parameters from light curves, 
such as the planet-to-star radius ratio. Planetary radii values are generally prone to such systematics.
However, apart from the presence of transits, no variability has been reported for HD 209458.
We stress the good consistency between the values of the planetary radius obtained from our stellar radius 
and the HST/STIS light curve in the optical, and the \textit{Spitzer}/IRAC light curve in the infrared. 
We conclude that the determination of the radius of the planet is not affected by this issue.

As previously mentioned the parallax uncertainty is the main contributor to that in the stellar radius. 
If using the modeled value $\Pi$=21.12$\pm$0.72 mas by \citet{torres2008}
the stellar radius decreases to $R_\star$=1.15$\pm$0.04~R$_{\sun}$. 
From this value and our determination of $M_\star$ using {\sevensize PARSEC} (see \S\ref{isochrones}), we found 
$\rho_\star$=1.05$\pm$0.12~g~cm$^{-3}$ and $\log g$= 4.37$\pm$0.03~($g$ in cm~s$^{-2}$),
instead of the values obtained from the first approach presented in \S\ref{star:mgd}, which are
$\rho_\star$=0.91$\pm$0.11~g~cm$^{-3}$ and $\log g$= 4.33$\pm$0.04~($g$ in cm~s$^{-2}$).
Regarding the second approach described in \S\ref{star:mgd}, 
the new value of $R_\star$ has a small impact on $\log g$ (4.36 instead of 4.38; with $g$ in cm~s$^{-2}$), 
but it reduces $M_\star$ from 1.26 to 1.10 M$_{\sun}$. 
Note the latter is closer to the mass derived from {\sevensize PARSEC} models in \S\ref{isochrones}. 
Concerning $R_\text{p}$, it drops by $\sim$5\% to 1.35$\pm$0.05 $R_\text{J}$ 
when using the parallax of \citet{torres2008}. $a$ also drops, from 0.0490$\pm$0.0020 au to 0.0467$\pm$0.0016 au.
The latter is closer to the value of $a$=0.0471$\pm$0.0005 au 
from \citet{torres2008}\footnote{Note that from the values for $\frac{a}{R_\star}$ and $R_\star$ of these authors 
we find $a$=0.0470$^{+0.0006}_{-0.0007}$ au.}.
Finally, $M_\text{p}$ = 0.67$\pm$0.05 $M_\text{J}$ when using the parallax of \citet{torres2008}.
This value is similar to the dynamical mass of 0.64$\pm$0.09 $M_\text{J}$ obtained by \citet{snellen2010}.

Our results confirm that the radius of HD 209458b is too large for the composition and age of its host star, 
challenging current theory models of internal structure. 
\textit{Gaia} will provide a reliable parallax measurement, which can be used to improve on
the stellar and planetary parameters, in particular the stellar radius, with an expected precision better than 1\%.
This will be useful to check the goodness of the stellar evolutionary models through the comparison with
direct and nearly model-independent values.

\section{Conclusions}
\label{conclusions}

Absolute flux spectrophotometry can efficiently provide nearly model independent 
stellar angular diameters and effective temperatures for a broad range of spectral types (A-type and sun-like stars).
The values of the angular diameters derived from flux ratios are several times more precise than those from interferometry. 
Combining angular diameters with accurate parallaxes it is possible to obtain very accurate radii. 
For transiting systems, the high level of accuracy in the stellar radius can be translated to the planetary radius $R_\text{p}$
from the transit depth determination, and then to the parameters that are related to $R_\text{p}$.
We have illustrated the application of this methodology to the determination of fundamental parameters 
for the well-known star HD 209458 and its transiting exoplanet, obtaining 1$\sigma$ uncertainties of 0.3\% 
in $T_\text{eff}$, and 4\% in the radii of the star and its planet. 
The parallax to be provided by \textit{Gaia} will allow us to significantly improve the accuracy and precision in 
the stellar/planetary radii, which will be then limited by the precision in the stellar angular diameter (0.75\%).

\section*{Acknowledgements}

This work has been supported by Mexican CONACyT research grant CB-2012-183007.
CAP is thankful to the Spanish MINECO for support through grant AYA2014-56359-P. 
This research has made use of the SIMBAD database, operated at CDS, Strasbourg, France and NASA's Astrophysics Data System.












\bsp	
\label{lastpage}
\end{document}